\documentclass[twocolumn]{aastex631}

\shorttitle{Photometric Calibration of CMOS-based Data}
\shortauthors{Xiao et al.}
\graphicspath{{./}{figures/}}

\usepackage{subfigure}
\usepackage{amsmath}
\usepackage{multirow}
\usepackage{tabularx}

\begin{document}

\title{Calibration of CMOS-based Photometry to a Few Milli-magnitude Precision: \\The Case of the Mini-SiTian Array}

\correspondingauthor{Kai Xiao; Yang Huang; Haibo Yuan}
\email{xiaokai@ucas.ac.cn; huangyang@ucas.ac.cn; yuanhb@bnu.edu.cn}

\author[0000-0001-8424-1079]{Kai Xiao}
\affiliation{School of Astronomy and Space Science, University of Chinese Academy of Sciences, Beijing 100049, People's Republic of China}
\affiliation{Institute for Frontiers in Astronomy and Astrophysics, Beijing Normal University, Beijing, 102206, China}

\author[0000-0003-3250-2876]{Yang Huang}
\affiliation{School of Astronomy and Space Science, University of Chinese Academy of Sciences, Beijing 100049, People's Republic of China}
\affiliation{Key Laboratory of Optical Astronomy, National Astronomical Observatories, Chinese Academy of Sciences, Beijing 100101, China}

\author[0000-0003-2471-2363]{Haibo Yuan}
\affiliation{Institute for Frontiers in Astronomy and Astrophysics, Beijing Normal University, Beijing, 102206, China}
\affiliation{School of Physics and Astronomy, Beijing Normal University No.19, Xinjiekouwai St, Haidian District, Beijing, 100875, China}

\author[0000-0002-7598-9250]{Zhirui Li}
\affiliation{New Cornerstone Science Laboratory, National Astronomical Observatories CAS, Beijing 100101, People's Republic of China}
\affiliation{School of Astronomy and Space Science, University of Chinese Academy of Sciences, Beijing 100049, People's Republic of China}

\author[0000-0002-3935-2666]{Yongkang Sun}
\affiliation{Key Laboratory of Optical Astronomy, National Astronomical Observatories, Chinese Academy of Sciences, Beijing 100101, China}
\affiliation{School of Astronomy and Space Science, University of Chinese Academy of Sciences, Beijing 100049, People's Republic of China}

\author[0000-0003-4573-6233]{Timothy C. Beers}
\affiliation{Department of Physics and Astronomy and JINA Center for the Evolution of the Elements (JINA-CEE), University of Notre Dame, Notre Dame, IN 46556, USA}

\author[0000-0001-6139-7660]{Min He}  
\affiliation{Key Laboratory of Optical Astronomy, National Astronomical Observatories, Chinese Academy of Sciences, Beijing 100101, China}

\author[0000-0002-2874-2706]{Jifeng Liu}
\affiliation{Key Laboratory of Optical Astronomy, National Astronomical Observatories, Chinese Academy of Sciences, Beijing 100101, China}
\affiliation{School of Astronomy and Space Science, University of Chinese Academy of Sciences, Beijing 100049, People's Republic of China}
\affiliation{Institute for Frontiers in Astronomy and Astrophysics, Beijing Normal University, Beijing, 102206, China}

\author{Hong Wu}  
\affiliation{Key Laboratory of Optical Astronomy, National Astronomical Observatories, Chinese Academy of Sciences, Beijing 100101, China}

\author{Yongna Mao}
\affiliation{Key Laboratory of Optical Astronomy, National Astronomical Observatories, Chinese Academy of Sciences, Beijing 100101, China}

\author[0000-0003-2471-2363]{Bowen Huang}
\affiliation{Institute for Frontiers in Astronomy and Astrophysics, Beijing Normal University, Beijing, 102206, China}
\affiliation{School of Physics and Astronomy, Beijing Normal University No.19, Xinjiekouwai St, Haidian District, Beijing, 100875, China}

\author[0009-0003-1069-1482]{Mingyang Ma}
\affiliation{Institute for Frontiers in Astronomy and Astrophysics, Beijing Normal University, Beijing, 102206, China}
\affiliation{School of Physics and Astronomy, Beijing Normal University No.19, Xinjiekouwai St, Haidian District, Beijing, 100875, China}

\author[0000-0002-7598-9250]{Chuanjie Zheng}
\affiliation{Key Laboratory of Optical Astronomy, National Astronomical Observatories, Chinese Academy of Sciences, Beijing 100101, China}
\affiliation{School of Astronomy and Space Science, University of Chinese Academy of Sciences, Beijing 100049, People's Republic of China}

\author[0009-0007-5610-6495]{Hongrui Gu}
\affiliation{Key Laboratory of Optical Astronomy, National Astronomical Observatories, Chinese Academy of Sciences, Beijing 100101, China}
\affiliation{School of Astronomy and Space Science, University of Chinese Academy of Sciences, Beijing 100049, People's Republic of China}

\author[0009-0001-5324-2631]{Beichuan Wang}
\affiliation{Key Laboratory of Optical Astronomy, National Astronomical Observatories, Chinese Academy of Sciences, Beijing 100101, China}
\affiliation{School of Astronomy and Space Science, University of Chinese Academy of Sciences, Beijing 100049, People's Republic of China}

\author[0000-0002-9824-0461]{Lin Yang}
\affiliation{Department of Cyber Security, Beijing Electronic Science and Technology Institute, Beijing, 100070, China}

\author[0000-0003-3535-504X]{Shuai Xu}
\affiliation{Institute for Frontiers in Astronomy and Astrophysics, Beijing Normal University, Beijing, 102206, China}
\affiliation{School of Physics and Astronomy, Beijing Normal University No.19, Xinjiekouwai St, Haidian District, Beijing, 100875, China}

\begin{abstract}
We present a pioneering achievement in the high-precision photometric calibration of CMOS-based photometry, by application of the Gaia BP/RP (XP) spectra-based synthetic photometry (XPSP) method to the mini-SiTian array (MST) photometry.
Through 79 repeated observations of the $\texttt{f02}$ field on the night, we find good internal consistency in the calibrated MST $G_{\rm MST}$-band magnitudes for relatively bright stars, with a precision of about 4\,mmag for $G_{\rm MST}\sim 13$. Results from more than 30 different nights (over 3100 observations) further confirm this internal consistency, indicating that the 4\,mmag precision is stable and achievable over timescales of months. 
An independent external validation using spectroscopic data from the Large Sky Area Multi-Object Fiber Spectroscopic Telescope (LAMOST) DR10 and high-precision photometric data using CCDs from Gaia DR3 reveals a zero-point consistency better than 1\,mmag. Our results clearly demonstrate that CMOS photometry is on par with CCD photometry for high-precision results, highlighting the significant capabilities of CMOS cameras in astronomical observations, especially for large-scale telescope survey arrays.
\end{abstract}

\journalinfo{Accepted by ApJL on March 6, 2025}


\section{Introduction} \label{sec:intro}
Charge-coupled devices \citep[CCDs;][]{1970BSTJ...49..587B} and complementary metal-oxide-semiconductor sensors \citep[CMOS;][]{1997ITED...44.1689F} both use the photoelectric effect to record photons, making them suitable for optical astronomical observations. 
Recently, CCDs have been widely used in astronomical observations owing to their advantages, such as producing high-quality images and good stability. Notable examples include the Sloan Digital Sky Survey \citep[SDSS;][]{2000AJ....120.1579Y}, Pan-STARRS \citep[PS1;][]{2002SPIE.4836..154K}, the Javalambre Physics of the Accelerating Universe Astrophysical Survey (J-PAS; \citealt{2014arXiv1403.5237B}), the Dark Energy Survey \citep[DES;][]{2015AJ....150..150F,2016MNRAS.460.1270D}, Gaia \citep{2016A&A...595A...1G}, SkyMapper \citep{2018PASA...35...10W}, J-PLUS \citep{2019A&A...622A.176C}, S-PLUS \citep{2019MNRAS.489..241M}, and SAGES \citep{2023ApJS..268....9F}. These surveys have already made, or are poised to make, profoundly significant contributions to modern astronomy.

However, for large-scale survey arrays consisting of dozens or hundreds of telescopes, the use of CCD detectors leads to significant economic costs. In contrast, CMOS detectors are considerably more cost-effective. Nowadays, some survey projects have adopted CMOS detectors, such as the Asteroid Terrestrial-impact Last Alert System (ATLAS) project \citep{2018PASP..130f4505T}, the SiTian\footnote{``SiTian" is a term from Chinese, referring to a governmental department established as early as the Yuan Dynasty in China to oversee astronomical observations and the formulation of the calendar.} project (SiTian; \citealt{2021AnABC..93..628L}), and the Large Array Survey Telescope \citep[LAST;][]{2023PASP..135f5001O}, featuring high sampling rates and large fields of view.

There is evidence that CMOS detectors can match CCDs in terms of electronic performance, such as high linearity, low dark current, low readout noise, and the variations of the pixel-to-pixel response and gain, as demonstrated by studies such as \cite{zhang25} and \cite{xiao25}. They have both conducted extensive studies on the ZWO ASI6200MM Pro CMOS cameras. \cite{zhang25} performed laboratory tests that demonstrated the CMOS sensors' exceptional performance, with a low average dark current rate of approximately 0.002 $e^-$ $\cdot$ pixel$^{-1}$ $\cdot$ s$^{-1}$ at $0^{\circ}$, consistent nonlinearity within $\pm$ 0.5\% (and even better, below 0.3\%, for signals exceeding 3000 ADU), and a low median read noise of 1.028 $e^-$. Furthermore, \cite{xiao25} findings confirmed that pixel-to-pixel variation remains stable over the course of a year, with a consistency better than 0.1\%.
However, a more prevalent concern is whether CMOS-based photometry can achieve the photometric-calibration accuracy typically found for CCD-based measurements.
To some extent, this influences perceptions of the potential for CMOS detectors to replace CCDs in large-scale time-domain survey projects.

With the stellar parameters provided by large-scale spectroscopic surveys such as LAMOST \citep{2012RAA....12.1197C,2012RAA....12..723Z}, high-quality photometric data from Gaia, and the release of Gaia DR3 BP/RP (XP) spectra \citep{2021A&A...652A..86C,2023A&A...674A...1G}, we have already improved and mastered the technology of the stellar color regression method \citep[SCR;][]{2015ApJ...799..133Y} and the ``corrected'' Gaia XP \citep{2024ApJS..271...13H} spectra-based synthetic photometry method (XPSP; introduced in \citealt{2023A&A...674A..33G} and improved by \citealt{2023ApJS..269...58X}). The improved XPSP method uses ``corrected'' Gaia XP spectra \citep{2024ApJS..271...13H} and achieves higher precision in synthetic photometric standard stars. Multi-band magnitudes are directly derived from the corrected spectra, without the need for 343 spectral coefficients. These methods enable us to construct a large number of high-precision standard stars, which can help precisely correct complex systematic errors from the Earth's atmosphere and instrumental effects. Successful applications to large-scale surveys have achieved high-precision calibration, with zero-point internal precision of a few milli-magnitudes. For instance, the SDSS Stripe 82 catalog has been recalibrated to a precision of 2--5\,mmag for both colors \citep{2015ApJ...799..133Y} and magnitudes \citep{2022ApJS..259...26H}. Gaia DR2 and EDR3 photometry were recalibrated to 1\,mmag \citep{2021ApJ...909...48N, 2021ApJ...908L..24Y}. The zero-points of the $uv$-bands in SkyMapper DR2 were also recalibrated to better than 10\,mmag \citep{2021ApJ...907...68H}. The recalibration of PS1 achieved a precision of 1--2\,mmag \citep{2022AJ....163..185X, 2023ApJS..268...53X}. Similarly, SAGES $gri$-band photometry was calibrated to 1\,mmag precision \citep{xiao}. Moreover, the photometry of J-PLUS DR3, S-PLUS DR4, and USS DR1 was recalibrated to a zero-point precision of 1--6\,mmag \citep{2023ApJS..269...58X,2024ApJS..271...41X,2025arXiv250103513L}.

This letter, using the photometric data of mini-SiTian array (MST) as an example, aims to pioneer the achievement of uniform CMOS-based photometry with a precision of 1\,mmag in the zero points, thereby demonstrating the great potential of CMOS detectors in contemporary astronomical observations. The organization of this letter is as follows. We construct our standard-star sample in Section\,\ref{sect:lib}, followed by a description of the photometric calibration for MST data in Section\,\ref{sect:homo}. The results and discussion are presented in Section\,\ref{sect:disscussion}, and brief conclusions are provided in Section\,\ref{sect:conclusion}.

\section{Data}
\label{sect:lib}
\begin{figure}
\centering
\resizebox{\hsize}{!}{\includegraphics{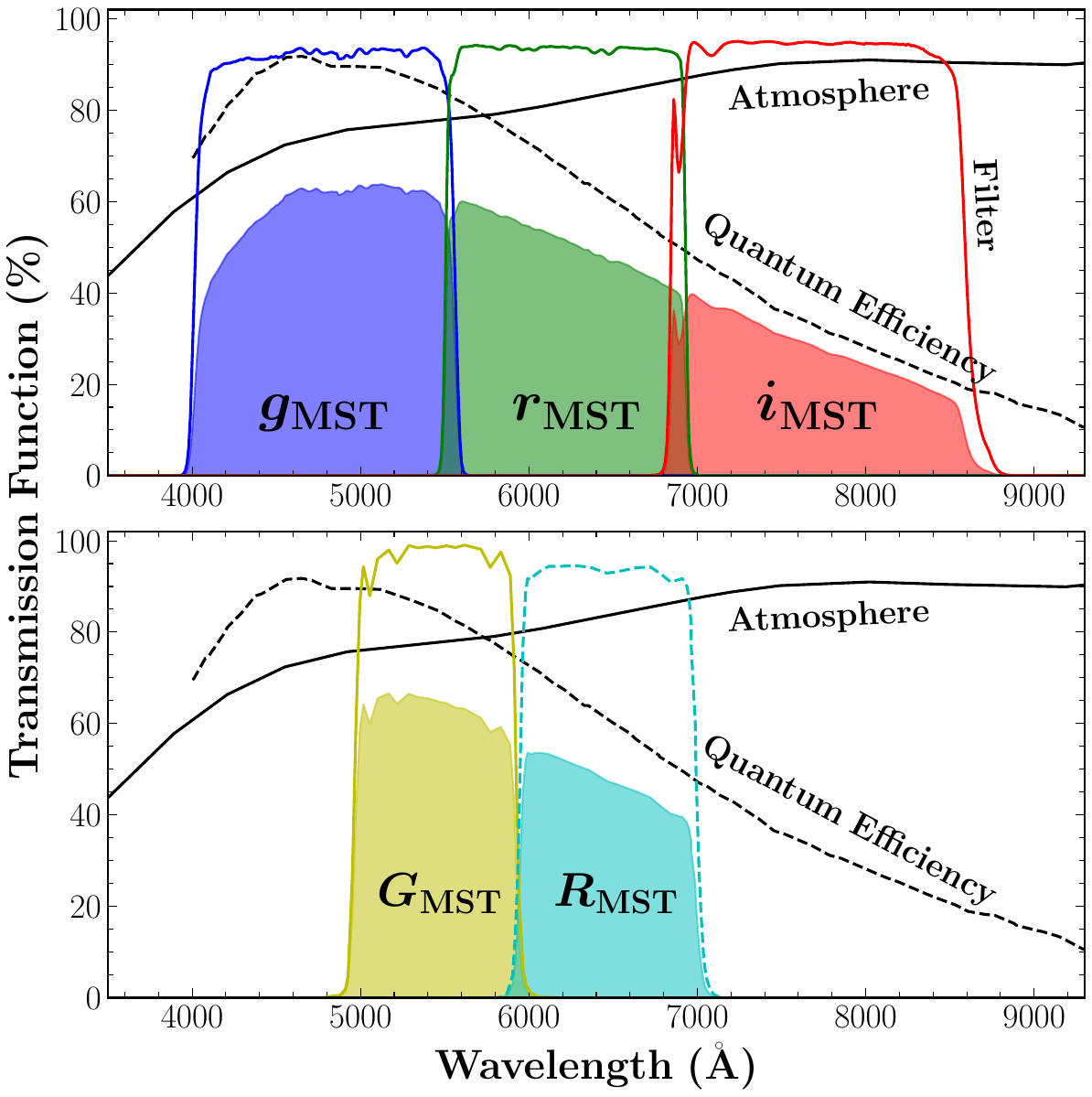}}
\caption{Synthetic transmission curves for the MST $G_{\rm MST}$ (yellow region), $R_{\rm MST}$ (cyan), $g_{\rm MST}$ (dark blue), $r_{\rm MST}$ (green), and $i_{\rm MST}$ (red) filters. The atmospheric transmission curve (black-solid curve), filter transmission curves (colored curves), and quantum efficiency (black-dotted curve) are also shown, along with the filter passbands. The influence of the telescope's optical system is not considered in this work. To clearly illustrate the shapes of the transmission curves across different passbands, the filter transmission curves have been divided into two panels.}
\label{Fig1}
\end{figure}
\subsection{MST Transmission Functions} \label{sec:21}
Due to the impact of the Earth's atmosphere, telescope optics, filters, and detector efficiency, not all light emitted by objects received at the top of Earth's atmosphere can be fully detected and recorded. The transmission function, $R_\chi(\lambda)$, for the $\chi$-band, incorporates components from the atmosphere, optical system, filter, and detector, and is written as:
\begin{eqnarray}
    \begin{aligned}
    R_\chi(\lambda) = R_{\rm atm}(\lambda) \cdot R_{\rm opt}(\lambda) \cdot R_{\rm filter}(\lambda) \cdot R_{\rm det}(\lambda). \label{tc}
    \end{aligned}
\end{eqnarray}
Here, $R_{\rm filter}(\lambda)$ and $R_{\rm det}(\lambda)$ represent the transmission function of the filters and the quantum efficiency of the detector, respectively, both of which are sourced from vendors. $R_{\rm atm}(\lambda)$ is defined as the atmospheric efficiency for an airmass of 1.2, computed as $10^{-0.4\times 1.2\times A(\lambda)}$, where $A(\lambda)$ is the typical atmospheric extinction values at Xinglong Station, obtained using the method described in \cite{2000PASP..112..691Y}. Here, the typical extinction curve does not account for certain absorption lines, such as the absorption in the A-band of oxygen from about 7580\,\AA~to 7780\,\AA, resulting in a subtle impact on the shape of the transmission curve. Additionally, the influence of the telescope's optical system, $R_{\rm opt}(\lambda)$, is negligible in this work. 
The optical system's transmission function and atmospheric absorption lines could slightly affect the overall transmission curve. For example, the optical system's efficiency varies by less than 10\% between 4000\,\AA~ and 10000\,\AA~\citep{Han25}. This may introduce a small color-dependent difference between the standard-star photometry (discussed in the next section) and the MST-observed photometry.
The resulting transmission curve for MST is shown in Figure\,\ref{Fig1}.

\subsection{The XPSP Standard Stars} \label{22}
By projecting the stellar spectral-energy distribution from the ``corrected'' Gaia XP spectra \citep{2024ApJS..271...13H} onto MST's transmission curves, we can integrate across the passbands and determine the synthetic magnitudes of stars observed with Gaia. 

Building upon previous works (e.g., \citealt{2012PASP..124..140B, 2023A&A...674A..33G}) and following the approach outlined in \cite{2023ApJS..269...58X,2024ApJS..271...41X}, we calculate the synthetic magnitude of over 200 million stars in the AB system \citep{1983ApJ...266..713O, 1996AJ....111.1748F} for the five MST bands ($G_{\rm XPSP}$, $R_{\rm XPSP}$, $g_{\rm XPSP}$, $r_{\rm XPSP}$, and $i_{\rm XPSP}$) using the ``corrected'' Gaia XP spectra-based XPSP method. 

These standard stars, which are brighter than approximately 17.65\,mag in the Gaia $G$-band, are distributed across the celestial sphere, typically ranging from a few thousand to tens of thousands per square degree, ensuring robust calibration coverage for the MST field.

\subsection{MST and \texttt{f02} field} \label{sec:23}
As outlined in \cite{2021AnABC..93..628L}, SiTian is a globally distributed network of 1-meter-class telescopes, strategically deployed across China and various international sites. Its primary scientific goals are the detection, identification, and monitoring of optical transients on previously unexplored timescales, typically under one day. To achieve these, SiTian will conduct high-cadence observations of the same sky, with intervals no longer than 30 minutes.

As a pathfinder of SiTian, the mini-SiTian (MST) array is located at the Xinglong Observatory and comprises three Schmidt telescopes (MST1, MST2, and MST3) with identical refractive optical configurations. The primary mirror of each telescope has a diameter of 30\,cm. To achieve a large field of view, the telescope is configured with a fast focal ratio of $f/3$ and equipped with a large-format CMOS camera, featuring the Sony IMX455 ZWO ASI6200MM Pro CMOS detector, positioned at the Cassegrain focus. The detector offers a resolution of 9k$\times$6k pixels, covering a 3 deg$^2$ field of view, at a pixel scale of $0.86^{\prime\prime}$ per pixel. 

\begin{figure}
\centering
\resizebox{\hsize}{!}{\includegraphics{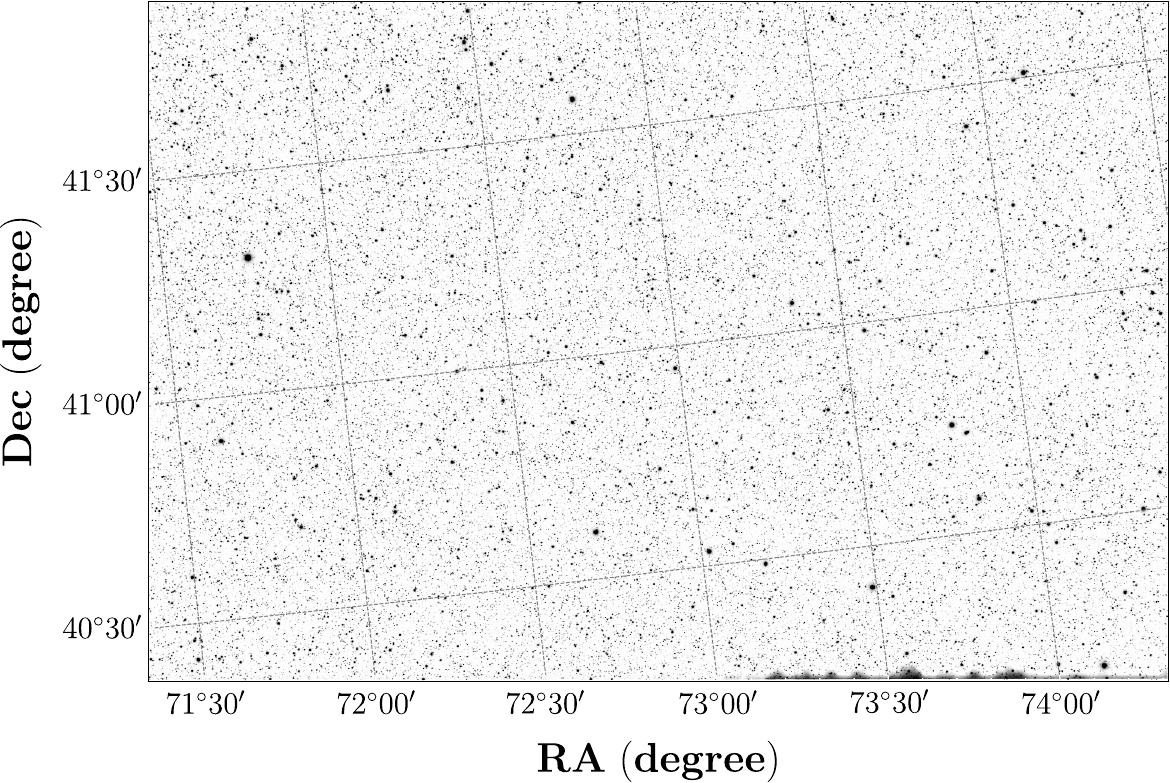}}
\caption{{The MST2 image of the test sky field, \texttt{f02}, has a center located approximately at $\text{RA} = 73^\circ$ and $\text{Dec} = 41^\circ$. The field contains roughly 50,000 stars.
}}
\label{Fig0}
\end{figure}

Since 2022, the MST has conducted extensive repeated observations of specific sky regions using 
$G_{\rm MST}/g_{\rm MST}$-, $R_{\rm MST}/r_{\rm MST}$-, $i_{\rm MST}$-filters installed on MST2, MST3, and MST1, respectively. 
For example, MST2, equipped with the $G_{\rm MST}$ filter, has collected over 3100 exposures of the \texttt{f02} test field (shown in Figure\,\ref{Fig0}), spanning more than 45 days, with each exposure lasting 300 seconds. In fact, the number of exposures varied on different days. Typically, more than 50 exposures were collected per night. For example, on January 18, 2023, 79 exposures were taken, while 92 exposures were collected on January 21, 2023.
The typical full width at half maximum (FWHM) of stars and the $5\sigma$ limiting magnitude are about $3^{\prime\prime}$ and 19.37\,mag. This test region, defined by $RA=73^\circ \pm 1.3^\circ$ and $Dec=41^\circ \pm 1.0^\circ$, is particularly suitable for calibration testing due to its low latitude ($|b| \sim 0.2^\circ$), ensuring that a large number of photometric standard stars can be used.

\subsection{MST Photometry}
\cite{xiao25} focuses primarily on the precise data processing pipeline designed for wide-field, CMOS-based devices, including the removal of instrumental effects, astrometry, photometry, and flux calibration. In the MST data processing pipeline, the flat-field correction for all images observed each night is performed using a master flat frame combined from the small-scale flats taken over the 11 closest nights within one month before and after this night. Aperture photometry is obtained using the mature software \texttt{SExtractor} \citep{1996A&AS..117..393B}, and we have set 25 evenly spaced aperture results ranging from 2 to 50 pixels. For the aperture correction, the pipeline first selects the aperture size with the highest signal-to-noise ratio (SNR) as the best aperture in each image, based on hundreds of bright, unsaturated sources. Then, the reference aperture for each image is determined using the growth curve of these same stars. Finally, a second-order two-dimensional polynomial, as a function of $X$ and $Y$ on the detector, is used to perform the aperture correction for each source in each image. 

Specific details of this process can be found in \cite{xiao25}. The MST aperture magnitudes after aperture correction are used as the default in this work.

\section{Method of Photometric Homogenization}
\label{sect:homo}
Photometric calibration is the process of correcting systematic errors caused by the Earth’s atmosphere and instruments. Those systematics can be expressed as a function of time, magnitude, color, and detector position $(X, Y)$, to ensure consistency in the observed magnitudes of celestial objects at different times, observing positions, and under varying observing conditions. To correct such systematics, for each MST image we combine MST data with Gaia DR3 photometry and the XPSP standard stars (refer to Section\,\ref{22}), using a cross-matching radius of $1^{\prime\prime}$. We then select calibration sample stars satisfying the constraints that magnitude errors are less than 0.01\,mag for MST $G_{\rm MST}$ band and $\texttt{SEXTractor~FLAG=0}$.

The photometric calibration is performed based on the calibration stars in each image using the following steps.
\begin{enumerate}
  \item[a.] Validation and correction of color- or magnitude-dependent systematic errors. When assessing the dependence of magnitude offsets, defined as $\Delta G=G_{\rm XPSP}-G_{\rm inst}$, on MST instrumental magnitude $G_{\rm inst}$ and Gaia DR3 color $G_{\rm BP}-G_{\rm RP}$, we observe a weak color-dependent (less than 0.01\,mag) but no magnitude-dependent systematics. The color-dependent systematic errors, potentially arising from differences between the XPSP and MST photometric systems (as discussed in Section\,\ref{sec:21}), are described using a second-order polynomial with three free parameters ($a_{0}$, $a_{1}$, and $a_{2}$) as a function of $G_{\rm BP}-G_{\rm RP}$. The polynomial is expressed as:
  \begin{equation}
    \Delta G(G_{\rm BP}-G_{\rm RP})=\sum_{i=0}^2 a_{i}\cdot (G_{\rm BP}-G_{\rm RP})^i~.
  \end{equation}
  The color-dependent systematics are then corrected by adjusting the XPSP standard magnitude as $G_{\rm XPSP}-\Delta G(G_{\rm BP}-G_{\rm RP})$. 

\begin{figure*}
\centering
\includegraphics[width=\textwidth, angle=0]{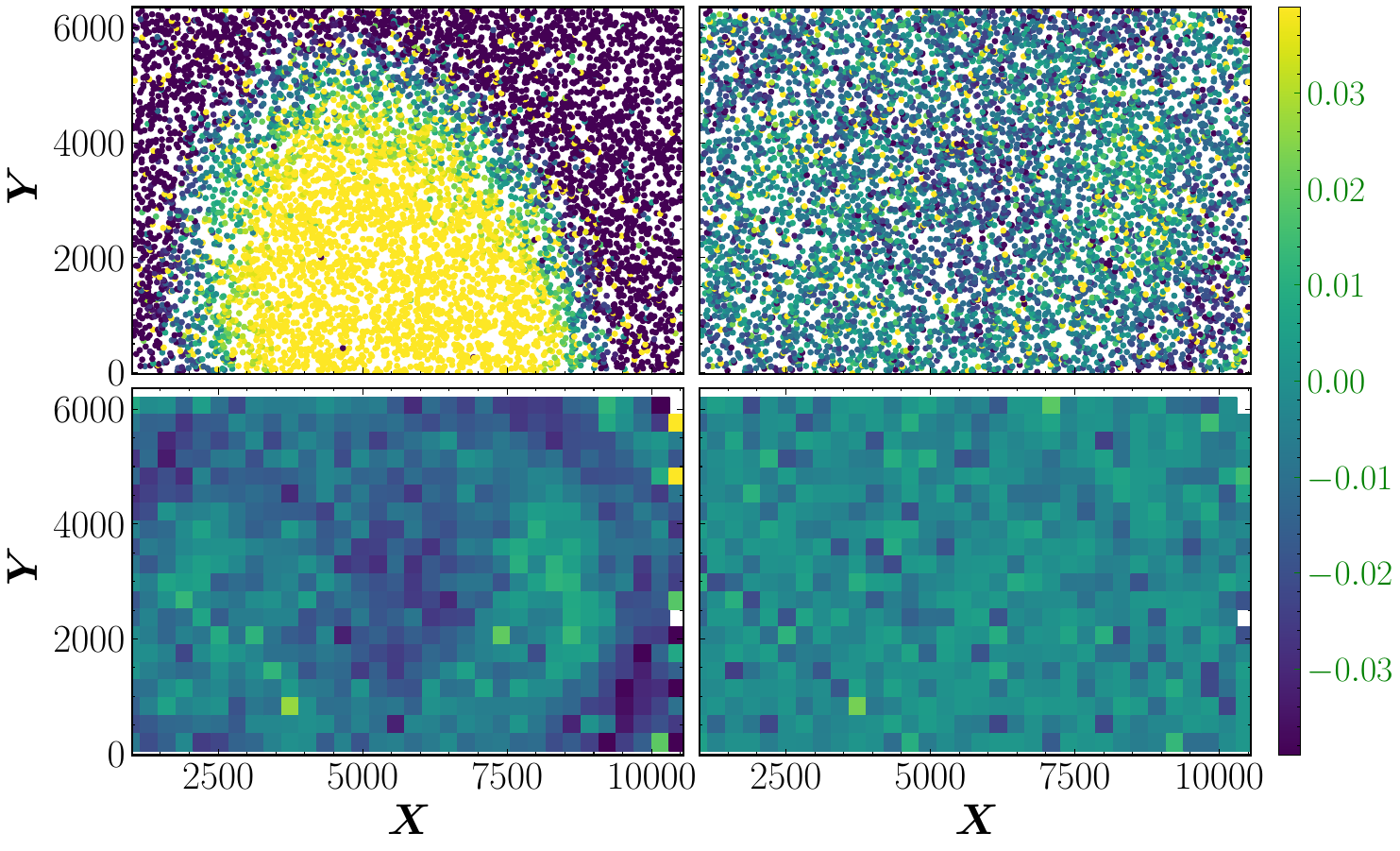}
\caption{Taking 20230118 as an example, showing the process of correcting spatial-dependent systematic errors. Upper-left panel: Distribution of the magnitude offsets after color-dependent systematic errors correction from the first observation. Upper-right panel: Distribution of the residuals after applying a second-order two-dimensional polynomial fitting to the magnitude offsets, which were previously corrected for the color-dependent systematics, from the same observation. Bottom-left panel: Distribution of the medium scale flat-field structure from all the 79 observations. Bottom-right panel: Same as the bottom-left panel, but after the correction for medium-scale flat-field structure. A color bar indicating the residuals is overplotted at the right.}
\label{Fig2}
\end{figure*}
  
  \item[b.] Validation and correction of spatial-dependent systematic errors. After color-dependent correction, we observe significant spatial structures, up to 0.05\,mag, in the dependence of magnitude offsets on spatial coordinates $(X, Y)$. To address this, a second-order two-dimensional polynomial with six free parameters is employed, as a function of $(X, Y)$, to fit the magnitude offsets: $G_{\rm XPSP}-G_{\rm inst}-\Delta G(G_{\rm BP}-G_{\rm RP})$. The polynomial model is given by:
  \begin{equation}
    \Delta G(X, Y) = \sum_{i=0}^{2} \sum_{j=0}^{i}b_{j, i-j} \cdot X^{j} \cdot Y^{i-j}~.
  \end{equation}
  
  To account for spatial-dependent errors, the corrected magnitudes are calculated as: $G_{\rm inst}+\Delta G(X, Y)$. The corrected magnitudes are then incorporated back into the previous step, and the process is iterated.
  
  \item[c.] The numerical stellar flat correction. After correcting for spatial-dependent errors, the calibrated MST $G_{\rm MST}$ magnitude is obtained as $G_{\rm MST} = G_{\rm inst}+\Delta G(X, Y)$.
  However, upon comparing the predicted magnitudes $G_{\rm XPSP}$ (after color correction) and the calibrated MST magnitudes $G_{\rm MST}$ in the $(X, Y)$-panel, a medium-scale spatial structure exceeding 0.01\,mag is observed. To correct for this, the numerical stellar flat correction method \citep{2024ApJS..271...41X} is employed. This method selects 20 standard stars around each star in the image for local linear fitting, yielding correction values denoted as $\delta G(X, Y)$, which are used to address these residual errors. This approach allows for the simultaneous determination of the flat-field correction for each source in the image. Finally, the calibrated MST magnitude $G_{\rm MST}$ is expressed as: 
  \begin{equation}
    G_{\rm MST}=G_{\rm inst}+\Delta G(X, Y)+\delta G(X, Y)~.
  \end{equation}
\end{enumerate}

Note that, despite adjusting the XPSP standard magnitude, the MST's absolute zero point remains effectively calibrated to the Gaia system due to the extremely weak color term.

\begin{figure*}
\centering
\includegraphics[width=\textwidth, angle=0]{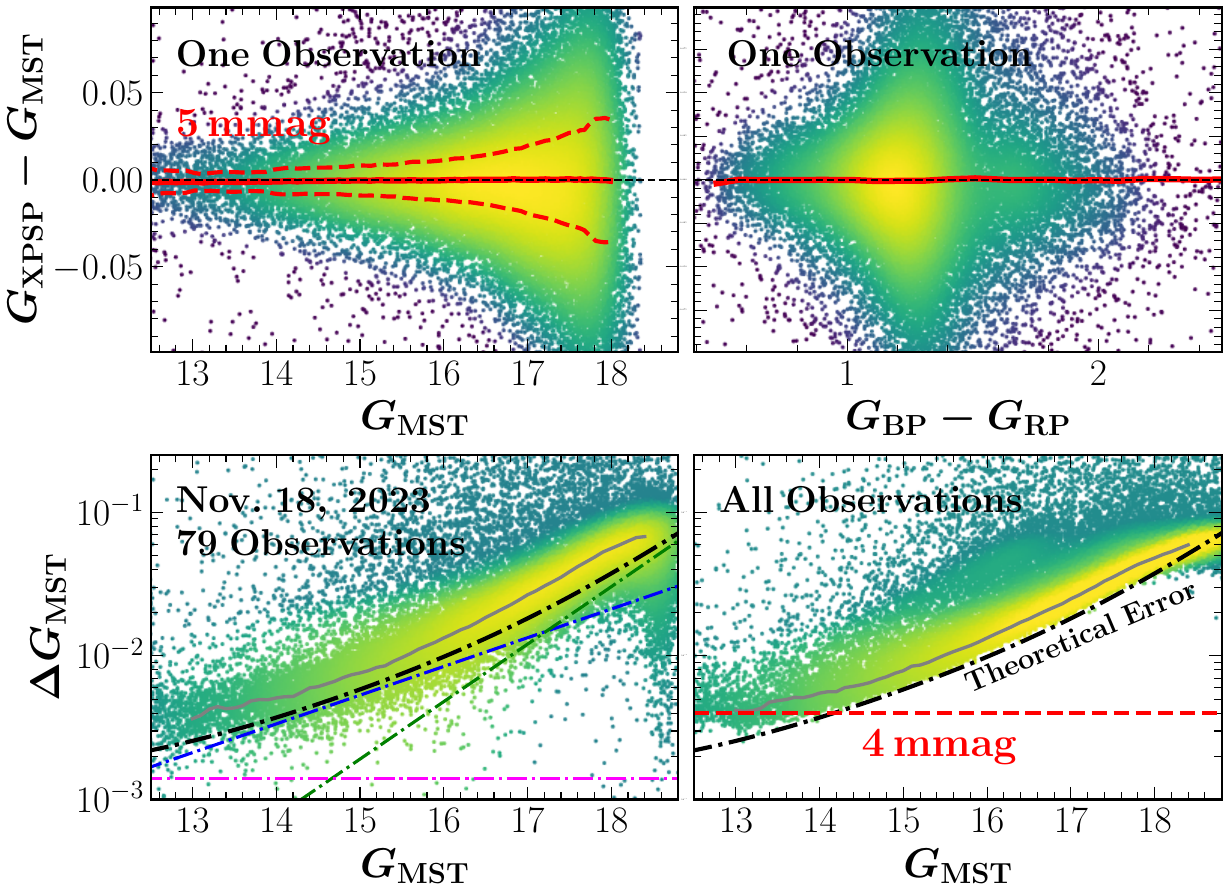}
\caption{Upper-left panel: The final fitting residuals against calibrated MST $G_{\rm MST}$ magnitude from one observation on January 18, 2023. Additionally, the panel includes red-solid and red-dotted curves representing the dependence of the median value and standard deviation on the MST-calibrated magnitude, respectively. Upper-right panel: Similar to the upper-left panel, but showing the dependence on Gaia color instead. Bottom-left panel: The variation in dispersion for 79 repeat observations of a common source on January 18, 2023, plotted against the calibrated magnitude $G_{\rm MST}$. The pink, blue, and green-dashed curves represent scintillation noise, photon noise, and sky noise, respectively. Bottom-right panel: Similar to the bottom-left 
panel, but includes over 3,100 observations from more than 30 different nights. The red-dotted lines represent a precision of 4\,mmag. The black-dashed curves and gray curves in the bottom panels represent theoretically calculated magnitude errors, as a function of magnitude and median values, respectively. }
\label{Fig3}
\end{figure*}

\begin{figure*}
\centering
\includegraphics[width=\textwidth, angle=0]{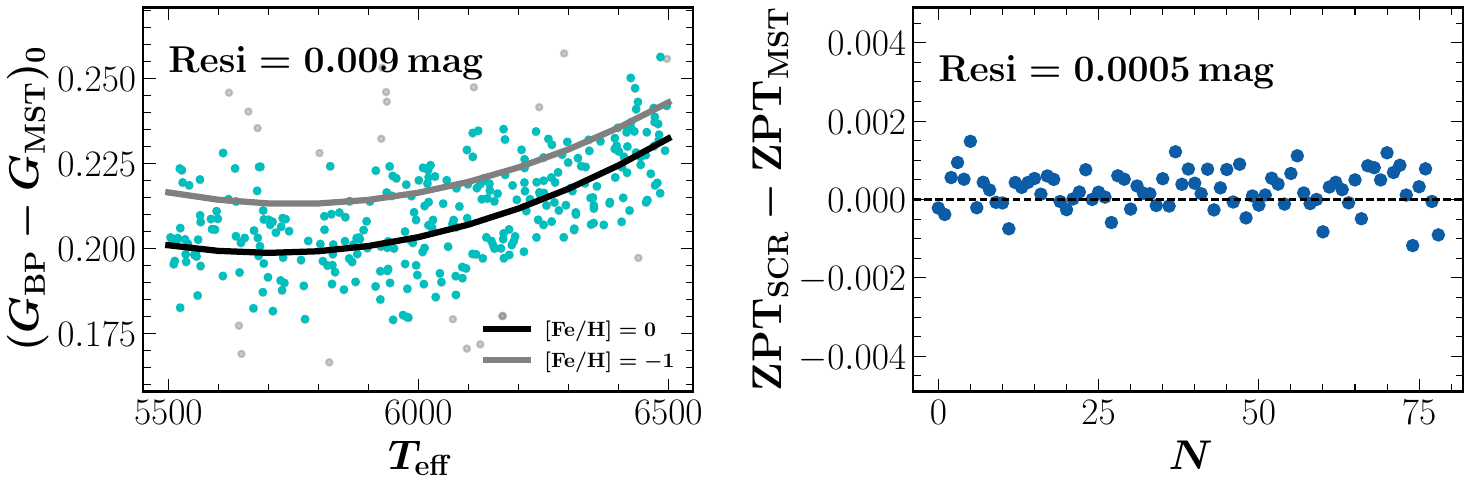}
\caption{Left panel: Two-dimensional second-order polynomial fitting (with six free parameters) of intrinsic color as functions of $T_{\rm eff}$ and $\rm [Fe/H]$, for the SCR calibration sample. The black and gray curves represent results for $\rm [Fe/H]=0$ and $-1$, respectively, and the standard deviation of the fitting residuals are estimated with Gaussian fitting and labeled in the panel. Right panel: Difference between the SCR zero-points and the MST zero-points. The zero-level is denoted by the black-dotted line, and the standard deviation of the residual is also provided.}
\label{Fig4}
\end{figure*}

\section{Result and Discussion}
\label{sect:disscussion}
As mentioned in Section\,\ref{sec:23}, the $\texttt{f02}$ field is highly suitable for testing calibration results. In this letter, we present the calibration results using more than 3,100 repeated observations.

First, using 79 images continuously exposed on January 18, 2023 as an example, we demonstrate the final correction results for spatial-dependent systematic errors. As shown in the upper-left panel of Figure\,\ref{Fig2}, spatial-dependent errors of up to 0.05\,mag are present, primarily due to systematic errors in the large-scale flat field correction. After applying second-order two-dimensional polynomial fitting, the residuals reveal a mottling structure, as illustrated in the upper-right panel of Figure\,\ref{Fig2}. Upon checking the 79 images taken that night, we found that the same structure was nearly consistent across all observations. To clearly display this medium-scale spatial structure, we combined the 79 residuals and performed binning into $30\times 30$ bins, as shown in the bottom-left panel of Figure\,\ref{Fig2}. This medium-scale structure, intermediate between large-scale and small-scale flat fields, is typical in modern photometric survey data, such as SAGES DR1, J-PLUS DR3, and S-PLUS DR4 \citep{xiao,2023ApJS..269...58X,2024ApJS..271...41X}. As depicted in the bottom-right panel of Figure\,\ref{Fig2}, the spatial distribution of the residuals became flatter in the $(X, Y)$-panel after applying a numerical stellar flat correction \citep{2024ApJS..271...41X}, reducing the standard deviation of the magnitude offsets from 5.1\,mmag to 1.1\,mmag.

To assess the consistency of our photometric calibration, we conducted both internal validations and external inspections.  As an example, for one observation on January 18, 2023, the differences between the standard magnitudes $G_{\rm XPSP}$ after color-term correction and the calibrated MST magnitudes $G_{\rm MST}$, as a function of $G_{\rm MST}$ or Gaia colors $G_{\rm BP}-G_{\rm RP}$, are shown in the upper panels of Figure\,\ref{Fig3}. These results demonstrate no dependence on MST magnitudes $G_{\rm MST}$ or $G_{\rm BP}-G_{\rm RP}$ colors.  From inspection of the upper-left panel of 
Figure\,\ref{Fig3}, the standard deviation of the differences increases at fainter magnitudes, as expected. At the bright end, the standard deviation is about 0.005\,mag at $G=13$, with more than 3000 observations of the $\texttt{f02}$ field exhibiting nearly identical results. 

The standard deviation of well-calibrated magnitudes from 79 observations of the $\texttt{f02}$ field and over 3000 observations of 30,000 stars, as a function of magnitude, is estimated with Gaussian fitting and plotted in the bottom panels of Figure\,\ref{Fig3}, along with the theoretical results of signal-to-noise ratio \citep[see][]{xiao25}. The deviation decreases at brighter magnitudes, aligning with theoretical predictions, and stabilizes at approximately 4\,mmag for $G<13$. It then rises slightly to around 0.01\,mag at $G\sim 15.5$, and increases to about 0.03\,mag towards the faint end at $G\sim 17$, stable across both single-night and multi-night observations. Our results suggest that the internal precision of photometric calibration using CMOS-based photometry can achieve a precision of 4\,mmag, comparable to that of CCD cameras.

\begin{figure*}
\centering
\includegraphics[width=\textwidth, angle=0]{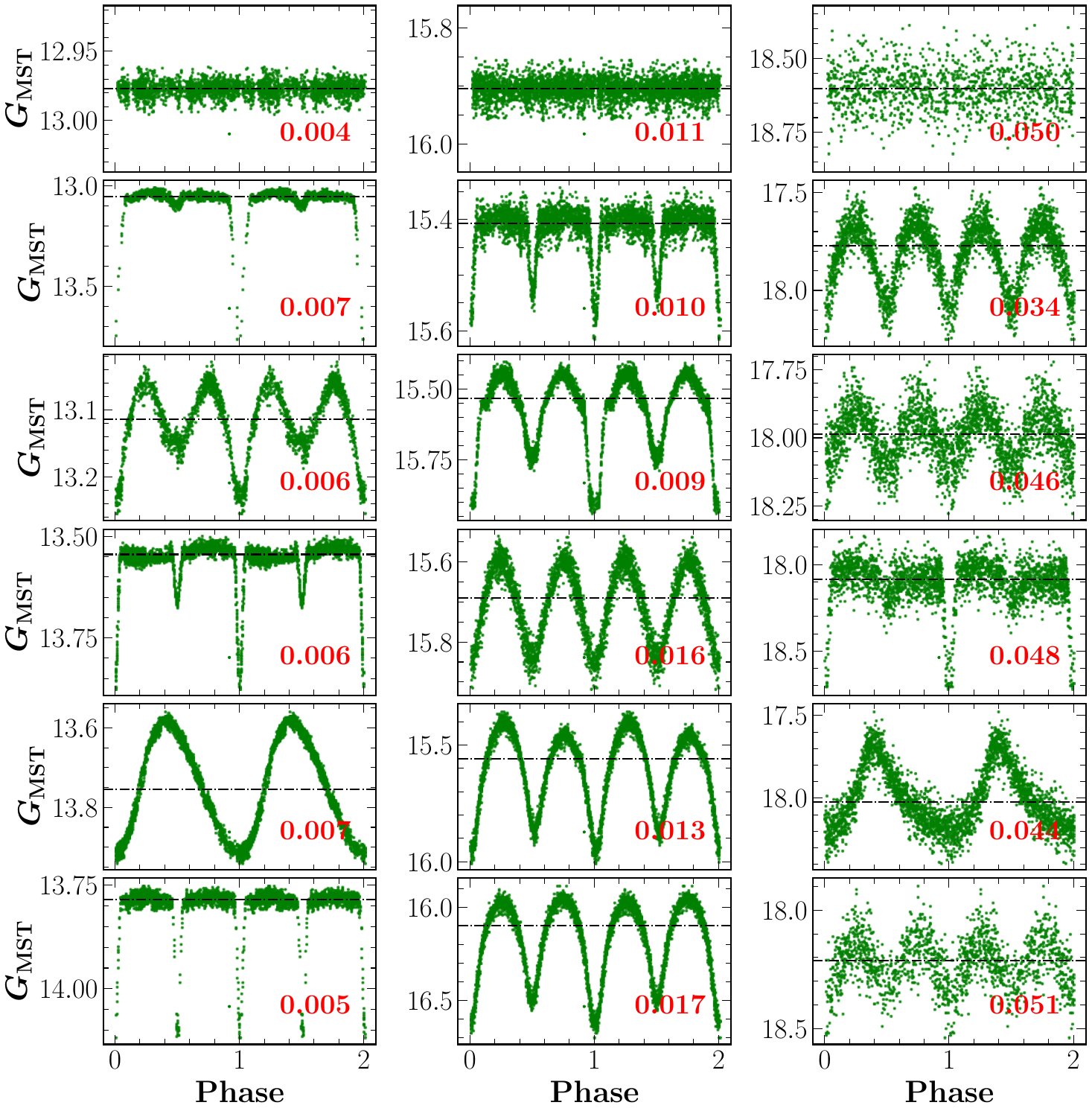}
\caption{Light curves from randomly selected stars—both non-variable (first row) and variable (other rows)—are presented based on over 3100 observations, illustrating varying brightness levels. Median values are plotted with black-dotted lines. For non-variable stars, the standard deviation of the magnitude is estimated using a Gaussian fitting approach and indicated in each panel. To analyze a variable source, we first obtain the median of the light curve, then linearly interpolate to derive model values for all points, and finally use Gaussian fitting to determine the standard deviation of the differences between model values and measured values. The standard deviation is labeled in the bottom-right corner of each panel.}
\label{Fig5}
\end{figure*}

As depicted in the bottom panels of Figure\,\ref{Fig3}, the trend of the data points aligns closely with the theoretical expectations. In this process, we accounted for the influence of scintillation noise \citep{1967AJ.....72..747Y,2015MNRAS.452.1707O} by calculating it for each image ($\sigma_i$), and then deriving the total scintillation noise ($\sqrt{\frac{1}{n}\sum_{i=1}^n \sigma_i^2}$). Here, $n$ represents the number of observations. We then computed the differences between the median values of the data points (the gray-solid curves) and the theoretical curve (the black-dashed curves). Our results reveal that, for the 79 \texttt{f02} observations and over 3,000 observations over a period of months, the consistency between the theoretical curve and the median value of the data points is 0.8\,mmag and 1.1\,mmag at $G_{\rm MST}\sim 13$, respectively.

To further validate the calibration precision of MST photometry, we employed an independent 
comparison. The first approach utilized the SCR method, combined with calibrated MST photometry from the initial observation on January 18, 2023, Gaia DR3 photometry, and spectroscopic data from LAMOST DR10. We selected main-sequence stars (${\log g}>-3.4\times 10^{-4}\times T_{\rm eff}+5.8$) as calibration stars, applying specific criteria: magnitude errors less than 0.01\,mag; $\texttt{phot}$\_$\texttt{bp}$\_$\texttt{rp}$\_$\texttt{excess}$\_$\texttt{factor}$ $<$ $1.3+0.06\times(G_{\rm BP}-G_{\rm RP})^2$ to exclude stars with problematic Gaia $G_{\rm BP}/G_{\rm RP}$ photometry; $5500<T_{\rm eff}<6500$\,K, and $\rm [Fe/H]>-1$ to ensure a constrained parameter space for robust fitting of intrinsic colors with atmospheric parameters while maintaining sufficient star counts; and a SNR for the $G_{\rm MST}$ band ($\rm SNR_{\rm g}$) in the LAMOST spectra exceeding 20. Ultimately, approximately 240 calibration stars were selected. 

To account for reddening, we adopted $E(G_{\rm BP}-G_{\rm RP})$ values derived from the star-pair method \citep{2013MNRAS.430.2188Y}. The reddening coefficient $R$ for $G_{\rm BP}-G_{\rm MST}$ color with respect to $E(G_{\rm BP}-G_{\rm RP})$, was modeled using a second-order two-dimensional polynomial based on $T_{\rm eff}$ and $E(G_{\rm BP}-G_{\rm RP})$: $R = 6.116\times 10^{-8}\times T_{\rm eff}^2 + 1.442\times E(G_{\rm BP}-G_{\rm RP})^2 - 4.051\times 10^{-4}\times T_{\rm eff} \times E(G_{\rm BP}-G_{\rm RP}) - 5.839\times 10^{-4} \times T_{\rm eff} + 1.265\times E(G_{\rm BP}-G_{\rm RP}) + 1.522$. Here, the reddening coefficient was determined iteratively using a sample of low-extinction stars and an initial value \citep{2023ApJS..264...14Z}. We then performed a fitting process to establish the intrinsic color as a function of $T_{\rm eff}$ and $\rm [Fe/H]$ using a two-dimensional polynomial. The fitting results, illustrating intrinsic colors as a function of $T_{\rm eff}$ and $\rm [Fe/H]$, are shown in Figure\,\ref{Fig4}. The standard deviation of the residual from intrinsic color fitting is about 9\,mmag, primarily dominated by the random errors in $T_{\rm eff}$ and $G_{\rm MST}$. For example, we calculated the typical error to be about 8.5\,mmag, when the typical error of $T_{\rm eff}$, $\rm [Fe/H]$, and $G_{\rm MST}$ are 10\,K, 0.02\,dex, and 6\,mmag, respectively.

Finally, we assessed the consistency of zero-points between the SCR method and calibrated MST photometry. We systematically selected $20\times 20$ evenly spaced points across the CMOS position space for each image. The median values of the zero-points from these 400 locations were adopted as the constant zero-point, as illustrated in the right panel of Figure\,\ref{Fig4}. The standard deviation of the zero-points between the SCR method and the MST zero-points consistently remains close to 0, with a scatter of approximately 0.5\,mmag.

To further visually illustrate this consistency, we selected 3 non-variable and 15 variable sources at different brightness levels. The light curves of these stars are shown in Figure\,\ref{Fig5}. Additionally, we quantitatively assessed the consistency of more than 3100 observations for non-variable sources using Gaussian fitting. We found the consistency to be approximately 4\,mmag at $G_{\rm MST}\sim 13$, 11\,mmag at $G_{\rm MST}\sim 16$, and 50\,mmag at $G_{\rm MST}\sim 18.5$. These results are consistent with those shown in the bottom-left panel of Figure\,\ref{Fig3}. For variable sources, we first fold the light curve into phase space using the known period. Next, we divide the phase space into 20 evenly spaced bins and calculate the median magnitude for each bin. By applying linear interpolation to the resulting median light curve, we subtract the light curve variations from each observed data point. Finally, we fit a Gaussian function to the residuals to derive the photometric errors from the light curves of the variable sources. These results align with the consistency observed for the non-variable sources.

% \vskip 3cm
\section{Conclusions}
\label{sect:conclusion}
In this study, we constructed the MST transmission function by accounting for atmospheric effects, filters, and detector efficiency. Using the improved XPSP method, we developed the MST photometric system's standard stars by convolving the transmission function with ``corrected'' Gaia DR3 XP spectra.

Based on these standards, high-precision calibration of MST $G_{\rm MST}$ photometry was achieved. We quantified color- and spatial-dependent systematic errors, identifying weak color dependencies and significant spatial variations. Color systematics, possibly due to transmission function inaccuracies, were corrected with a second-order one-dimensional polynomial. Large-scale systematics, arising primarily from flat-field errors, were addressed using a two-dimensional polynomial fit, while middle-scale errors were mitigated by a numerical stellar flat correction.

Internal validation, via 79 and more than 3000 repeat observations for the $\texttt{f02}$ field, revealed $G_{\rm MST}$ consistency of approximately 4\,mmag for $G<13$, about 10\,mmag for $G\sim 16$, and around 50\,mmag for $G\sim 18$. External validation, integrating MST data with LAMOST DR10 and Gaia DR3 photometry using the SCR method, yielded zero-point alignment with SCR that was better than 1\,mmag, confirming the Sony IMX455 ZWO ASI6200MM Pro CMOS's suitability for high-precision time-domain surveys.

Our results have clearly demonstrated that CMOS-based photometry has achieved parity with CCD-based photometry for high-precision measurements. 
However, the clustering of observed sources within 60 pixels may have underestimated small-scale flat-field effects, which warrants further investigation. Moreover, CMOS's independent pixel operation can introduce electronic instabilities (e.g., varying pixel gains) and  ``salt \& pepper'' noise \citep{2023PASP..135e5001A}, which impact faint-object detection. Once these problems are carefully quantified and addressed, CMOS detectors have the potential to offer significant advantages over CCD detectors for contemporary astronomical observations.

\newpage
% \begin{acknowledgments}
This work is supported by the National Natural Science Foundation of China grants No. 12403024, 12422303, 12222301, and 12173007; the National Key R\&D Program of China (grants Nos. 2023YFA1608300 and SQ2024YFA160006901); the Postdoctoral Fellowship Program of CPSF under Grant Number GZB20240731; the Young Data Scientist Project of the National Astronomical Data Center; and the China Postdoctoral Science Foundation No. 2023M743447.
T.C.B. acknowledges acknowledge partial support for this work from grant PHY 14-30152; Physics Frontier Center/JINA Center for the Evolution of the Elements (JINA-CEE), and OISE-1927130: The International Research Network for Nuclear Astrophysics (IReNA), awarded by the US National Science Foundation.  

The SiTian project is a next-generation, large-scale time-domain survey designed to build an array of 60 optical telescopes, primarily located at observatory sites in China. This array will enable single-exposure observations of the entire Northern sky with a cadence of only 30 minutes, capturing 
true-color ($gri$) time-series data down to about 21 mag. This project is proposed and led by the National Astronomical Observatories, Chinese Academy of Sciences (NAOC). As the pathfinder for the SiTian project, the Mini-Sitian project utilizes an array of three 30 cm telescopes to simulate a single node of the full SiTian array. The Mini-Sitian began its survey in November 2022. The SiTian and Mini-SiTian have been supported from the Strategic Pioneer Program of the Astronomy Large-Scale Scientific Facility, Chinese Academy of Sciences and the Science and Education Integration Funding of University of Chinese Academy of Sciences.

This work has made use of data from the European Space Agency (ESA) mission
Gaia (\url{https://www.cosmos.esa.int/gaia}), processed by the Gaia
Data Processing and Analysis Consortium (DPAC,
\url{https://www.cosmos.esa.int/web/gaia/dpac/consortium}). Funding for the DPAC
has been provided by national institutions, in particular the institutions
participating in the Gaia Multilateral Agreement.
Guoshoujing Telescope (the Large Sky Area Multi-Object Fiber Spectroscopic Telescope LAMOST) is a National Major Scientific Project built by the Chinese Academy of Sciences. Funding for the project has been provided by the National Development and Reform Commission. LAMOST is operated and managed by the National Astronomical Observatories, Chinese Academy of Sciences.

\software{Astropy \citep{2022ApJ...935..167A}, Matplotlib \citep{2007CSE.....9...90H}, NumPy \citep{2020Natur.585..357H}, SciPy \citep{2020NatMe..17..261V}}
% \end{acknowledgments}

% \clearpage
% \section{Appendix}
% For F02 field, we have over 40 days of observations. From the daily data, we select one image each day. We then assess the consistency of brightness measurements for the same source across approximately 40 repeated observations, as shown in Figure\,\ref{Fig:A1}.

% \begin{figure}[ht!] \centering
% \subfigure{\includegraphics[width=9.4cm]{f6.pdf}} \\
% \caption{{\small Same as the bottom left panel of Figure\,\ref{Fig3}, but from different observation day.}}
% \label{Fig:A1}
% \end{figure}

\end{document}